# On Improving Research Methodology Course at Blekinge Institute of Technology

Shoaib Bakhtyar[1] and Ahmad Nauman Ghazi[2]
[1]shoaib.bakhtyar@gmail.com, [2]nauman.ghazi@bth.se
[1]Department of Computer Science and Engineering
[2]Department of Software Engineering
Blekinge Institute of Technology, Karlskrona, Sweden.

## Abstract

The *Research Methodology in Software Engineering and Computer Science* (RM) is a compulsory course that must be studied by graduate students at *Blekinge Institute of Technology* (BTH) prior to undertaking their theses work. The course is focused on teaching research methods and techniques for data collection and analysis in the fields of Computer Science and Software Engineering. It is intended that the course should help students in practically applying appropriate research methods in different courses (in addition to the RM course) including their Master's theses. However, it is believed that there exist deficiencies in the course due to which the course implementation (learning and assessment activities) as well as the performance of different participants (students, teachers, and evaluators) are affected negatively. In this article our aim is to investigate potential deficiencies in the RM course at BTH in order to provide a concrete evidence on the deficiencies faced by students, evaluators, and teachers in the course. Additionally, we suggest recommendations for resolving the identified deficiencies. Our findings gathered through semi-structured interviews with students, teachers, and evaluators in the course are presented in this article. By identifying a total of twenty-one deficiencies from different perspectives, we found that there exist critical deficiencies at different levels within the course. Furthermore, in order to overcome the identified deficiencies, we suggest seven recommendations that may be implemented at different levels within the course and the study program. Our suggested recommendations, if implemented, will help in resolving deficiencies in the course, which may lead to achieving an improved teaching and learning in the RM course at BTH.

## 1. Introduction

In many Swedish universities students are required to study a mandatory course concerning research methods. Different Swedish universities, such as [1, 2, 3, 4], offer a course that focuses on research methods in order to train students in appropriate research practices and methods that may be used in their research. At *Blekinge Institute of Technology* (BTH), *Research Methodology in Software Engineering and Computer Science* (RM) [3] is one such course, offered to Master's degree students in Computer Science (CS) and Software Engineering (SE) prior to undertaking their theses work. The course is equivalent to 7.5 credits (in the European Credit Transfer and Accumulation System) and it is offered in fourth Learning Period (LP) to Swedish students who are already enrolled in CS and SE study programs at BTH. The first LP at BTH begins in September each year and the fourth LP begins in March the following year. Every year, in January, international exchange students from different countries (such as, India and China) join Master's in CS and SE programs at BTH. Therefore, the fourth LP (for Swedish students) is in practice second LP for international exchange students.

The course is focused on teaching of research methods, and techniques for data collection and analysis in the fields of CS and SE. The course is designed based on the principle of outcome-based education [5], i.e., it is clearly stated in the course descriptor what a student should know or should be able to do after successfully completing the course. Students' knowledge in the course is evaluated using three assignments. The first assignment is a *systematic literature review* (SLR), where students have to conduct a SLR on a topic of interest. In the second assignment, students are given the task of writing a research proposal based on the research gaps identified in their SLR and their future plans for the selected topic. In the third assignment, the students are expected to write a research paper for their selected topic. In all of the assignments, it is expected from the students to apply the knowledge learned within the course on a research topic of interest.

The overall purpose of the RM course is to guide students through the process of conducting a research project from beginning to end. The course is intended to help students in practically applying appropriate research methods not only in the RM course but also in other courses including their Master's theses. However, it is found that the students are skeptical about the role of RM course when conducting their theses [6]. A study by Shahid & Shahzad [6] investigates problems related to thesis topic selection by graduate students in a multicultural educational environment at BTH. Through a survey and focus group discussions with international students at BTH, the authors in [6] found that international students find it challenging to select a thesis topic at BTH although there exists well defined routine for thesis topic selection. Furthermore, it is found that there exist several problems related to the RM course [6]. The participants of the study had mixed opinions about the course, e.g., the participants were unsure about the role of RM and whether the course is useful for strengthening their research skills. Another interesting response was provided by participants with backgrounds in CS and SE. CS students considered the course to be less useful as compared to SE students. The authors in [6] suggest to revise the curriculum of RM course. However, no concrete evidence is provided concerning exact deficiencies in the course that need to be resolved.

In addition to the study by Shahid & Shahzad [6], there have also been discussions over the years among different teachers and evaluators involved in the RM course. The course teachers and evaluators agree that there exist deficiencies in the course. However, there is a lack of clear evidence to substantiate these claims. An example of a potential deficiency, where both the evaluators and teachers agree, is that CS and SE are two different departments at BTH and the course is offered to a mix audience from both the departments, which results in teaching and evaluating students with different backgrounds. The problem intensifies when CS students are evaluated by SE teachers and vice-versa.

The purpose of this article is to identify deficiencies in teaching and learning in the RM course at BTH and to suggest recommendations for overcoming the deficiencies. Using a case study approach, we aim to identify the deficiencies that exist in teaching and learning the RM course at BTH. We believe that the identified deficiencies might be resolved; therefore, in addition to identifying deficiencies in the course, our aim is also to suggest guidelines for overcoming the identified deficiencies. In addition to providing concrete evidence of the deficiencies in the RM course, our goal is to suggest recommendations that may be practically applicable in the RM course at BTH in order to improve the teaching and learning experience in the course. The knowledge gained in the course is potentially used in research, i.e., Master's degree thesis, and hence by improving the learning experience of students in the course we expect to achieve improvement in students' theses quality. Furthermore, we argue that identifying the deficiencies and proposing recommendations to overcome those deficiencies (both from teaching and learning perspectives) will help in achieving the intended learning outcome of the course in an improved manner.

The remaining of this article is structured as follows: Section 2 presents the research methodology followed in this article. Section 3 presents a discussion on the results of our study and our recommendations on how to overcome the identified deficiencies. Finally, in Section 4, the article is concluded by presenting our concluding remarks and pointers to the future.

## 2. Methodology

The research methodology followed in this article is a case study research. Case studies are used to investigate a phenomenon in its natural settings that facilitates the researchers to gain an in-depth understanding of the case and the context [7]. In order to conduct this case study, we follow the guidelines for case study research in Software Engineering as suggested by Runeson and Höst [7].

### 2.1 Case study design

*Objectives:* The aim of this research study is to identify the deficiencies in teaching and learning activities within the RM course. Moreover, based on the knowledge gained through this exploratory case study, the long-term objective is to suggest recommendations on how to address the identified deficiencies, which may help the teachers in designing and teaching an effective RM course based on intended learning outcomes.

Overall goal of this study is formulated following the Goal-Question-Metric approach: To gain an in-depth understanding of the deficiencies (Purpose) in teaching and learning (Issue) of RM course in SE and CS (Object) both from teachers and students' perspective (Viewpoint).

### 2.2 Research questions

In this exploratory case study, we intend to answer the following research questions:

*RQ1: What are the key deficiencies in teaching and learning research methodology course?*
*RQ2: How can the identified deficiencies be resolved?*

### 2.3 Case and context

In this research study, the case has been identified as the RM course designed for students undertaking their master's degrees in SE and CS. The course is designed to introduce students to research methods that they shall use to conduct research within different courses as well as in their theses. Moreover, it is intended that the course should help in building a solid knowledge base of the students, thus, enabling the students to apply appropriate research methods for solving a particular research problem. Students are also encouraged during the course to learn necessary tools and skills that enables them to collect and analyze data gathered through the application of different research methods.

### 2.4 Data collection

We use semi-structured interviews as our data collection method. To select the interviewees, we have used a convenience sampling approach [8]. Our interviewees participated in the RM course in different time periods during the years 2008-2016 and in various roles, i.e., students, assignment evaluators (evaluator), and course responsible (instructor). In total, we interviewed 4 evaluators, 2 course responsible and 4 students.

#### 2.4.1 Questionnaire design

The questionnaire was designed considering multiple themes; such as course descriptor (question 1 and 2), overall course structure (questions 3-7), course administration (questions 8-10), students' participation in the course (questions 11-15), knowledge application from the course (question 16 and 17), and feedback provided in the course (question 18 and 19), which are central to the planning, execution, administration, and application of the course.

## 2.5 Data analysis

The interviews were recorded and later transcribed. To analyze the qualitative data collected from the interviews, we used thematic analysis technique [9]. The transcribed interviews were color-coded for data extraction. Different color codes were assigned to parts of the interviews for building different themes in order to answer our research questions stated in Section 2.2.

# 3. Results and discussion

Below, we present a discussion on the results from our interviews. The results are discussed from the perspectives of teachers (i.e., evaluators and course responsible) and students.

## 3.1 Course deficiencies from teachers' perspective

From the teachers' perspective, the following deficiencies exist in the RM course:

D1. We identified that the evaluation rubrics used for evaluating Master's thesis proposals are followed for evaluating assignments in the RM course. Since the rubrics are not designed by the teacher, a critical deficiency is how to interpret the rubrics. At present, we found that the course teacher interprets the rubrics according to the best of his/her knowledge and perception.

D2. Several rubrics are found to be overlapping and, hence, sometimes the same assignment content submitted by a student may be assessed multiple times, which often result in a student getting good marks if a specific part in his/her assignment was done better and vice versa.

D3. Some of the evaluators found the rubric scale to be confusing. For example, some evaluators are unsure about the difference between good and very good in the rubrics. This problem becomes more visible when a particular assignment is graded by two different evaluators, which results in the student receiving two completely different reviews for the same assignment.

D4. As mentioned earlier, the course has three assignments where in assignment 1 students are expected to report a SLR and in assignment 3 students perform and report an empirical study. It is noticed that these two assignments narrow down students' understanding on acceptable research methods in SE and CS. Many students view these two methods as the only options when working on their theses. This results in students being good at identifying problems but unable to discuss different research methods, i.e., having no deeper understanding of the research methods.

D5. Some evaluators view the description for SLR (assignment 1) to be not in line with the other guidelines by the course responsible. As there is no clear format given to the students to follow in assignment 1, the description for SLR is sometimes interpreted by students as a format for writing the assignment report. Furthermore, the allotted time for all the three assignments is considered to be unrealistic.

D6. Most often, the number of CS evaluators is less as compared to SE in the course, which results in some CS assignments to be evaluated by SE evaluators. This leads in SE evaluator taking more time to evaluate an assignment and most often the evaluators do not feel confident about their comments. From evaluators' perspective, having both CS and SE students is definitely a problem. A possible reason for this problem is because in SE department (in general) one is trained as an empirical researcher, whereas CS department follows a rationalistic approach in research.

D7. The allotted budget (time) for assignments' assessment is another problem. The course responsible agrees that it is not possible for one person to check each and every assignment and therefore evaluators are needed to solve this challenge. However, it is found that sometimes the assignment and the allotted budget time for assessment does not fit since some assignments need more time than the allocated time. For example, at one particular instance of the course in 2015, it was calculated that there were a total of 324 students reports to grade in the course which accounts to a total of 1620 pages. The evaluators found it challenging to read and provide extensive feedback on 1620 pages. The assessment task becomes even more challenging when half of these pages (approximately 800) are written by students from CS and evaluated by SE evaluators.

D8. Several deficiencies were identified regarding the placement of course, i.e., when the course must be offered to students. The course is offered in consultation with CS department and they view the course as a preparation for the Master's thesis. In contrast to CS department, course responsible at SE department believes that the earlier you take this course; the more benefits you will have in other courses within the study program, i.e., what is learned in the RM course can be practiced in other SE courses. This deficiency can be characterized as a curriculum knowledge challenge since both CS and SE follow different curriculums. Curriculum knowledge enables a teacher to perceive the relationship of a course in a particular study program [10] and it helps a teacher to better design the course in relation to the study program.

D9. The evaluators have mixed opinions on when the course must be offered to the students. Some evaluators do not recommend the course to be offered just immediately before the thesis. These evaluators suggest that the best time would be third LP (for international students) because in the first or second LP new students arrive in Sweden (with no or less knowledge about the culture norms and learning environment at BTH) and the RM course is an important course. On successful completion of the RM course, the students can apply the knowledge gained in the RM course in their other courses where they are required to perform an empirical study.

D10. Some evaluators experienced that the students are immature from research perspective in order to participate in the course. These evaluators believe that perhaps the student are participating in the course too early in their study program and that these students are unclear about scientific writing, plagiarism, and referencing. This leads to another complex problem of students asking for more detailed feedback on their assignments from the evaluators.

D11. During several instances of the course execution, it was noticed that international students join the course late (arriving late into Sweden) due to which they often miss important lectures that are aimed at guiding students for their assignments. The main reason for late joining is because most of the participants in the course are international students and these students often face delays when getting their visa for Sweden. It is estimated that on average if there are approximately 100 students in the course, 40 of them would be international students. This means 40% of the class participants miss the first few lectures and this affects their performance in the first assignment.

D12. During lectures, the teachers mostly experienced one-way communication, i.e., little or no students' participation during the lectures. According to one of the teacher, the reason for low participation was partly because of 100-150 students in a single class. The teacher believe that students' participation would have been different if the class consisted of a smaller number of participants.

D13. The evaluators and teachers were asked if the students use knowledge gained within the RM course in any other way outside the course. While the teacher agreed that students use some of the knowledge gained within the RM course in their final theses, the evaluators had different opinion. Some of the evaluators found, when supervising students in their theses, that the students have no knowledge about appropriate research methods even though the students had successfully completed the RM course.

**3.2 Course deficiencies from students' perspective**

From the perspective of students, the following deficiencies are identified in the RM course:

D14. The course descriptor is an important document that outlines key information about a course, such as what will a student learn in a course, which text books will be followed in a course, and how will the students be assessed. It was interesting for us to find that most of the interviewed students did not read the course descriptor when registering for the course. Most of the students were unaware about the course descriptor and its importance when they registered for the course. Each participant had a different reason for not reading the course descriptor. For example, one student claimed that he was unable to find the descriptor anywhere, neither did he know that he was supposed to read the descriptor. He agreed that, due to this, he had no idea what he will be studying and what is required to do in the course. Another student (who also did not read the course descriptor) claimed that he knew about the course contents and requirements from former students in the RM course.

D15. Late arrival in Sweden due to visa process was found to be another problem. Due to late arrival the students miss most of the lectures in beginning to the course. Due to the late arrival the students' performance was affected and they were unable to perform even basic tasks, such as using library databases for literature search. Furthermore, the basic concepts needed for assignments are covered in lectures at the course beginning and the students missed some of the lectures due to late arrival.

D16. During lectures, the students found it a steep learning curve to absorb contents from the course since the course contents and teaching method was considered to be quite different from the students' background. Most of the students had difficulty in finding a suitable topic for research in the course and even if the students find something, they would be unsure about their choice.

D17. Most of the students agreed that they could not apply all the taught concepts from the course in their assignments because some of the key concepts were not discussed in detail during the lectures. For example, data analysis techniques were briefly stated during the lectures and not discussed in details but these techniques are expected in the assignments. Furthermore, the students found the lectures to be very much theoretical, which made the lectures to be boring for participation. Additionally, some students agreed that they did not gain anything from the lectures because of their mental ability at that particular time when the course was offered to them. The students claimed that they could not even differentiate between basic concepts in the course, such as the difference between qualitative and quantitative methods. The main reasons for this problem was students' mental maturity (aptitude level) when they studied the course, as well as the lectures material which (according to the students) did not guide them properly.

D18. Regarding students' participation in lectures, the students agreed that they were reluctant to ask questions due to different reasons. Some students did not ask questions because of low self-confidence and if the students would be confused over a specific issue, they would later ask questions from their group members. Some students did not ask any questions during lectures due to a different culture background, where it is considered impolite to question the teacher.

D19. Most of the interviewed students agree that the course must be placed at the end of their study program due to different reasons. Some students believe that the course is really important for their Masters' thesis and if the course is offered at the end of the program then a late arriving student would not suffer in the course as well as in the thesis. Whereas, some of the international students claimed that when they studied the course in the second LP, they performed well in the course. However, by the time these students reached their Master's thesis, they could not recall some key concepts from the course that were needed in their thesis.

D20. Most of the interviewed students had no problems with the contents requirement for assignments. However, regarding the timeline for completing assignments there are mixed views. Some students, who had no prior knowledge about research, considered the timeline for completing assignments to be unrealistic. The students needed more time for the first (SLR) and third (empirical work) assignment. Other students claimed that only the first two assignments (SLR and Research Proposal) are enough for completion in a single LP.

D21. In the course, some of the CS students performed SLR by following the guidelines from the field of SE. Later (when these students successfully completed the course), they conducted SLR in their Master's thesis by strictly following the guidelines from the course. These students received critical comments on their SLR from theses examiners, who were from CS. This convinced the CS students that whatever they learned from an SE perspective may not be necessarily followed in CS, which means the students had a narrow understanding on the application of research methods during the course.

To summarize the deficiencies identified from our interviews, we use key components of a typical constructively aligned course as suggested by Biggs & Tang in [11]. In a constructive aligned course, the following key components are clearly defined and aligned [11]:

- Intended Learning Outcomes (ILOs),
- Teaching and Learning Activities (TLAs), and
- Assessment Tasks (ATs).

ILOs outline what the students will learn in a course [11]. In RM course descriptor [3], ILOs are clearly stated; however, we observed that some learning outcomes are too broad and not clearly defined. Examples of too broad ILOs from the RM course descriptor are:

- *Students should be able to discuss and relate to the concept of science and relate to it in their own work.*
- *Students should be able to discuss scientific possibilities, the role of knowledge in society, people's responsibility for how knowledge is used and the ethical and societal implications of a research project can bring.*

Similarly, some ILOs from the course descriptor, which are highly important for executing a research project, are not covered in the lectures. For example, the results from our interviews (with students) indicate that the following critical learning outcome is not discussed in detail in lectures:

- *Students should be able to describe different methods of research, data collection and analysis.*

We argue that there is a need to redesign and make the ILOs more concrete in the course descriptor. Additionally, there need to be measures taken for students to read the course descriptors because a major problem found is that the students are unaware of the role and importance of course descriptors prior to courses registration.

TLAs state what learning activities need to implemented in order to achieve the ILOs [11]. We have identified several deficiencies concerning TLAs in the RM course. Examples include: low students' participation, students arriving late to the course, students' inability to find a suitable topic for their research etc.

ATs define how well the students have achieved the ILOs [11]. In the RM course, the ATs are the three assignments (i.e., SLR, Research Proposal, and Empirical Work). We found several deficiencies in the ATs of the course. The deficiencies include: assessment rubrics are not realistic, the timeline for completing assignments is short, SE evaluators assessing CS reports and vice-versa, assignments missed by students due to late arrival, the allocated time for evaluators to check assignments is short etc.

### 3.3 Our recommendations

In order to overcome the identified deficiencies and improve the teaching and learning activities in the course, we herein suggest some recommendations that may be implemented at the course and study program level.

R1. A critical deficiency found was that most of the students do not read the course descriptor prior to enrolling in the course. The most common reason found is that the students are either unaware of the course descriptors or they do not know where to find the descriptors. Therefore, we suggest to ensure that the newly enrolled students are better informed and well aware about the role and importance of course descriptors. This can be done in several ways, e.g., important information about the role and importance of course descriptors may be stated in the admission letter or it may be provided as a supplement to the admission letter. Alternatively, the students can be informed about course descriptors in their orientation lecture in the very beginning of the study program.

R2. A deficiency, where both the evaluators and students agree, is the maturity (aptitude level) of students in the course. It can be because either the students have different background or they join the course late due to visa problems. This leads to poor performance by the students and extra evaluation time for the evaluators. To overcome this deficiency, we suggest that research topics should be given to the students in advance by the teacher(s). Alternatively, the teacher may provide abstract ideas to students and then the students can come up with a concrete topic based on that idea. This will make the students more confident that they are doing the right work and it will reduce the overhead for both the teachers and evaluators. Another alternatively can be to provide a common topic to the students and they can then be given the task to address the topic using different research methods that are taught in the course.

R3. At BTH, we found that the assessment tasks for another course (i.e., the Advanced Topic in Computing) are similar to the assignment 1 and 2 in the RM course. The course Advanced Topic in Computing [12] helps students develop skills in information retrieval, literature reviewing and scientific writing within topics of their interest, in preparation for their future master thesis projects. The course requires students to perform and report a SLR and a research proposal similar to assignment 1 and 2 in the RM course. Our recommendation is that there must be two assignments in the RM course. Assignment 1 can be about different research methods that can be used to solve a problem. This assignment will help the student to identify and analyze different research methods that can be used to solve a problem. Assignment 2 can be the same as it is today, i.e., empirical work by selecting a method from the discussed methods in assignment 1. The SLR and proposal assignment may be handled in the Advanced Topic in Computing course and that course may be made compulsory for students to attend.

R4. The assessment rubrics for evaluating assignments should be redesigned since (at present) the rubrics evaluating Master's thesis proposals are used in the course. Our recommendation here is that the rubrics should be redesigned in consultation with evaluators who have been involved in the course before. We believe, the evaluators (who have the experience of using the rubrics to evaluate assignments) know exactly the problems that exist in rubrics and the criteria that need to be added or removed from the rubrics.

R5. To address the problem of late arriving students and to ensure that the students are familiar with plagiarism, scientific writing etc., we recommend to offer the course in middle of the study program, i.e., in the third or fourth LP to international students. This recommendation will also help to address the evaluators' concern to offer the course in the third LP, which will help the students to apply the knowledge gained in the course in other courses.

R6. Students' participation can be improved by holding debates among students' groups. For example, students' groups can be given a topic and then there can be debates among the groups on different research methods that can be applied to solve the given topic. This recommendation will also make the lectures more interesting for students to participate actively.

R7. Finally, we recommend that the course should be offered separately to CS and SE students. This will solve different problems that exist today. By offering the course separately to CS and SE students, it will reduce the number of participants in the class, which will result in better management of students by the teacher/evaluators and increased participation from the students. It will also solve the problem of SE evaluators evaluating CS students and vice versa.

## 4. Concluding remarks

In this article, we investigated deficiencies concerning teaching and learning in the RM course offered to graduate students at BTH. Using semi-structured interviews with 2 course responsible (teachers), 4 evaluators, and 4 students, we found that there exist some critical deficiencies at different levels in the course. For example, at the course registration level, students are unaware of the course descriptor for the course. Additionally, the course is offered to a mixed audience of CS and SE, which makes it a challenging work for the evaluators to evaluate the assignments that are from a field other than the evaluator's field of research.

Furthermore, at different levels in the course, different participants (i.e., teachers, evaluators, and students) either have different opinions about the same problem or they are affected in different ways by the same problem. For example, from the view point of teacher, a reason for low participation by students in the lecture is due to a large class population. In contrast, from the students' perspective, they are reluctant to ask questions during the lecture due to a different cultural background or low self-confidence. Another such example is of placing the course in the very beginning of the study program. For the students, it is a problem because they miss some important lectures in the beginning of the course due to late arrival. For the evaluators, late arrival of students leads to the problem of handling assignments from students who are not fully equipped with knowledge needed for their assignments.

In order to resolve the identified deficiencies, we suggested different recommendations in Section 3.3. We claim that our suggested recommendations, if implemented, will help in resolving the deficiencies in the course, which may lead to achieving an improved teaching and learning experience within the RM course at BTH. We conclude that a deficiency may be resolved by implementing one or more alternative recommendations. For example, the deficiency *D9* may be resolved by either implementing recommendation *R5* or *R3*. Similarly, some deficiencies require a combination of multiple recommendations to be implemented, e.g., to resolve *D13*, implementation of *R2* and *R3* is needed. It is worth mentioning here that some of our suggested recommendations, if implemented, will require changes to be made to the course descriptor of the RM course. In particular, implementing recommendations *R3* and *R6* will require changes to be made to the course descriptor. *R3* suggest changes to the ATs in the RM course, i.e., the assignments. Whereas, *R6* suggest changes to the TLAs, i.e., holding debates among students' groups to increase students' participation and learning in the RM course.

|  | Recommendations | | | | | | |
|---|---|---|---|---|---|---|---|
| **Deficiencies** | R1 | R2 | R3 | R4 | R5 | R6 | R7 |
| D1 |  |  |  | X |  |  |  |
| D2 |  |  |  | X |  |  |  |
| D3 |  |  |  | X |  |  |  |
| D4 |  | X | X |  |  |  |  |
| D5 |  | X | X |  |  |  |  |
| D6 |  |  |  |  |  |  | X |
| D7 |  |  | X |  |  |  | X |
| D8 |  |  |  |  |  |  | X |
| D9 |  |  |  |  | X |  | X |
| D10 |  | X |  |  | X |  |  |
| D11 |  |  |  |  | X |  |  |
| D12 |  | X |  |  |  | X |  |
| D13 |  | X | X |  |  |  |  |
| D14 | X |  |  |  |  |  |  |
| D15 |  |  |  |  | X |  |  |
| D16 |  | X |  |  |  | X |  |
| D17 | X |  | X |  |  | X |  |
| D18 |  | X |  |  |  | X |  |
| D19 |  |  | X |  | X |  |  |
| D20 |  |  | X |  |  |  |  |
| D21 |  |  |  |  |  |  | X |

Table 1. Deficiencies & recommendations mapping.

In Table 1, we present a mapping of the identified deficiencies in the RM course and our recommendations. From the table, we can conclude that approximately more than half of the identified deficiencies (i.e., 11 out of 21) in the course (see, Section 3) can be resolved by our recommendation *R2* and *R3*. In our recommendation *R2*, we suggest that a concrete or abstract research topic should be given to the students in the beginning of the course. The students can then refine the given research topic according to their own needs and later they can apply different research methods on the refined topic. This recommendation will help resolve deficiencies *D4, D5, D10, D12, D13, D16* and *D18*. In *R3*, we suggest that there should be 2 assignments in the RM course because the first and second assignments in the course already exist in another course, i.e., the Advance Topic in Computer Science [12]. For assignment 1 in the RM course, we recommend analysis of different research methods. This will make the course more focused on different research methods that can be applied, which will result in students achieving a deeper understanding of the research methods. In assignment 2, the students can then select an appropriate research method from the first assignment and apply it on their research topic. *R3* will help in resolving deficiencies *D4, D5, D7, D13, D17, D19* and *D20*.

At present, the course responsible for RM course is in the process of redesigning the course. We, therefore, discussed our recommendations with him for the course and he has agreed to use most of our suggested recommendations. However, a possible limitation of our study is that we do not consider the budget factor (such as, cost, time, and other resources) when suggesting our recommendations. Therefore, future work includes a quantitative analysis of our suggested recommendations, i.e., a cost-benefit analysis when implementing our recommendations at BTH. We believe different resources, such as time and financial cost, may be affected when implementing our recommendations, while at the same time the recommendations may help in achieving improved teaching and learning experience in the course.

# Appendix - Interview Questions

1. What problems you noticed/identified in the course plan when you took the course?
2. Did you find any problems with the new course plan after it has been redesigned?
3. Do you find any problems with the lectures and assignments in the course?
4. Do you find any problems with the number of instructors in the course?
5. Do you think the time allotted for the whole course is enough?
6. Did you have any background of research methodologies prior to the course?
7. What do you think when should the course be offered (at the end of the program, at the beginning, or in the middle)?
8. What problems normally occur during course registration?
9. Is timeline for the tasks (assignments) realistic?
10. What is your experience/opinion about CS and SE students/teachers mismatch in the course?
11. What is your opinion about students' participation in the course (i.e., is it enough or less than expected)?
12. How can we improve students' participation in the lectures?
13. When it comes to interacting with the students, are the students reluctant or open to ask questions?
14. Have you found out any reason why the students are reluctant to ask questions?
15. What is the level of feedback on teaching from the students?
16. How do you (as a student) apply the knowledge gained through the course?
17. Do you see any evidence of students applying the knowledge gained in the course?
18. Why students do not perform as expected in the course?
19. Was the feedback given to the students utilized by them as expected?
20. If the feedback was not utilized by students as expected, can you explain what can be the possible reasons?
21. If the feedback was utilized by students as expected, how effectively do the students utilize the given feedback?